\documentclass[12pt]{article}

\InputIfFileExists{Debug}{}{}

\usepackage[utf8]{inputenc}
\usepackage{amsmath}
\usepackage{amsthm}
\usepackage{amsfonts}          % if you want the fonts
\usepackage{amssymb}           % if you want extra symbols
\usepackage[mathscr]{eucal}
\usepackage{graphicx}
\usepackage{color}
\usepackage{hypbmsec}
\usepackage{fancyvrb}
\usepackage{sepnum}
\usepackage{xspace}

\usepackage{rotating}
\usepackage{slashbox}
\usepackage[isu,bf]{caption}
\newlength{\xtrawidth}
\newlength{\xtraheight}
\usepackage[all]{xy}           % Commutative diagrams

\ifx\NoBackreferences\EMPTYMACRO
  \usepackage[backref,linktocpage,bookmarks,breaklinks]{hyperref}
\else
  \usepackage[linktocpage,bookmarks,breaklinks]{hyperref}
\fi
\usepackage{breakurl}

\ifx\DEBUG\EMPTYMACRO
\else
  \usepackage{showlabels}
\fi

\def\clap#1{\hbox to 0pt{\hss#1\hss}}

\def\mathrlap{\mathpalette\mathrlapinternal}
\def\mathclap{\mathpalette\mathclapinternal}

\def\mathrlapinternal#1#2{%
\rlap{$\mathsurround=0pt#1{#2}$}}
\def\mathclapinternal#1#2{%
\clap{$\mathsurround=0pt#1{#2}$}}	

\makeatletter
  \def\adots{\mathinner{\mkern2mu\raise\p@\hbox{.}
      \mkern2mu\raise4\p@\hbox{.}\mkern1mu
      \raise7\p@\vbox{\kern7\p@\hbox{.}}\mkern1mu}}
\makeatother

\newcommand{\eqdef}{%
 \mathrel{\lower.1mm
   \hbox{$\stackrel{\lower.424ex\hbox{\scriptsize def}}{=}$}}
}

\newcommand{\C}{\ensuremath{{\mathbb{C}}}}

\newcommand{\Z}{\mathbb{Z}}
\newcommand{\CP}{\ensuremath{\mathop{\null {\mathbb{P}}}}\nolimits}

\DeclareMathOperator{\Id}{id}
\DeclareMathOperator{\rank}{rank}

\DeclareMathOperator{\Sym}{Sym}

\DeclareMathOperator{\img}{img}
\DeclareMathOperator{\coker}{coker}
\DeclareMathOperator{\Ind}{Ind}

\DeclareMathOperator{\Hom}{Hom}

\DeclareMathOperator{\diag}{diag}
\DeclareMathOperator{\Vect}{Vect}

\newcommand{\Rep}[1]{\ensuremath{\mathbf{\underline{#1}}}}
\newcommand{\barRep}[1]{\ensuremath{{\overline{\Rep{#1}}}}}

\newcommand{\ZZZ}{\ensuremath{{\Z_5\times\Z_5}}}

\newcommand{\Osheaf}{\ensuremath{\mathscr{O}}}

\newcommand{\Ksheaf}{\ensuremath{\mathscr{K}}}

\newcommand{\dual}{\ensuremath{\vee}}

\newcommand{\Singular}{\textsc{Singular}\xspace}

\newcommand{\cyclperm}{\ensuremath{(\text{cyc})}}

{\end{list}}

{\everymath{\displaystyle\everymath{}}\array}%
{\endarray}

\begin{document}
%%%%%%%%%%%%%%%%%%%%%[ Title Page ]%%%%%%%%%%%%%%%%%%%%%%%%%%
\begin{titlepage}
  \vspace*{-2cm}
  \hfill
  \parbox[c]{5cm}{
    \begin{flushright}
%      arXiv:yymm.nnnn [hep-th]
%      \\
      DIAS-STP 09-08
    \end{flushright}
  }
  \vspace*{2cm}
  \begin{center}
    \Huge 
    Three Generations \\
    on the Quintic Quotient
  \end{center}
  \vspace*{8mm}
  \begin{center}
    \begin{minipage}{\textwidth}
      \begin{center}
        \sc 
        Volker Braun
      \end{center}
      \begin{center}
        \textit{
          Dublin Institute for Advanced Studies\hphantom{${}^1$}\\
          10 Burlington Road\\
          Dublin 4, Ireland
        }
      \end{center}
      \begin{center}
        \texttt{Email: vbraun@stp.dias.ie}
      \end{center}
    \end{minipage}
  \end{center}
  \vspace*{\stretch1}
  \begin{abstract}
    A three-generation $SU(5)$ GUT, that is $3 \times
    (\Rep{10}+\barRep{5})$ and a single $\Rep{5}$--$\barRep{5}$ pair,
    is constructed by compactification of the $E_8$ heterotic
    string. The base manifold is the $\ZZZ$-quotient of the quintic,
    and the vector bundle is the quotient of a positive monad. The
    group action on the monad and its bundle-valued cohomology is
    discussed in detail, including topological restrictions on the
    existence of equivariant structures. This model and a single
    $\Z_5$ quotient are the complete list of three generation
    quotients of positive monads on the quintic.
  \end{abstract}
  \vspace*{\stretch1}
\end{titlepage}
\tableofcontents

\section{Introduction}
\label{sec:Intro}

Monad bundles~\cite{Witten:1993yc, Distler:1993mk, Blumenhagen:1996vu,
  Anderson:2007nc, Anderson:2008uw, Anderson:2008ex, Guffin:2008pi}
are the largest known class of $(0,2)$-compactifications. However, so
far only monad bundles on simply-connected Calabi-Yau manifolds were
explicitly constructed. However, just as in the heterotic standard
embedding, free quotients are, amongst many other aspects, important
in reducing the particle spectrum. For example, the net number of
generations was found~\cite{Anderson:2008uw} to peak somewhere around
$60$. By dividing out the free action of a discrete group $G$, the
number of generations would be divided by $|G|$.

In this paper, I will construct a slope-stable rank $5$ vector bundle
on the $\ZZZ$-quotient of the quintic via the monad construction. The
best way to work with non-simply connected manifolds is, as usual, to
construct everything on the universal cover (the quintic). However,
care has to be taken to make everything symmetric under the group
action, and as we will see this imposes purely topological
restrictions on the Chern classes of the constituents of the
monad. Via the so-called ``non-standard embedding'', this bundle of
vanishing first Chern class defines a $(0,2)$-compactification of the
$E_8$ heterotic string, giving rise to a low-energy $SU(2)$ gauge
group. The resulting matter spectrum will be three generations of
$\Rep{10}+\barRep{5}$ together with an (optional) vector-like pair of
$\Rep{5}$--$\barRep{5}$. A hopefully useful \Singular worksheet
demonstrating the spectrum computation is in
\autoref{sec:singular}. Although $\Z_5$ Wilson lines cannot be used to
break $SU(5)$ to the standard model gauge group, a mechanism like
$U(1)_Y$-flux could conceivably be employed. In any case, the
technology for constructing equivariant monads will certainly be
useful for more thorough searches for heterotic standard models.

\section{The Quintic Quotient}
\label{sec:quintic}

The quintic $Q$ is the simplest Calabi-Yau manifold. As the name
suggests, it is given by the zero set of a sufficiently generic
degree-$5$ hypersurface $Q(z_0,z_1,z_2,z_3,z_4)=0$ in projective space
$\CP^4$. Using this description of a smooth hypersurface, one can show
that there are precisely two\footnote{The $\Z_5$ is a subgroup of
  $\ZZZ$, so there is one \emph{maximal} free group action.}  possible
free group actions on the quintic: $\Z_5$ and $\ZZZ$. However, the
complex structure of $Q$, that is, the quintic polynomial $Q(z)$, has
to be chosen suitably to admit this symmetry. Without symmetry, the
quintic has $h^{2,1}(Q)=101$ complex structure moduli. Imposing the
symmetries restricts the complex structure moduli to a $25$ and
$5$-dimensional stratum, respectively.

For the purposes of this paper, I will focus solely on the free $\ZZZ$
group action~\cite{Candelas:1985en}. It acts projectively linear on the homogeneous
coordinates of the ambient projective space; The two group generators
are ($\zeta = e^{\frac{2\pi i}{5}}$)
\begin{equation}
  \label{eq:gzact}
  g_1(z_i) = z_{i+1}
  ,\qquad
  g_2(z_i) = \zeta^i z_i  
  .
\end{equation}
They satisfy
\begin{equation}
  \label{eq:g1g2commutator}
  g_1^5=1=g_2^5
  ,\quad
  g_1 g_2 = \zeta^{-1} g_2 g_1
\end{equation}
and, therefore, define a $\ZZZ$ group action on $\CP^4$. If the
quintic is invariant, then this action defines a group action on the
hypersurface, too. The invariant polynomials are best described by a
Hironaka decomposition
\begin{equation}
  \label{eq:HironakaZ5Z5}
  \C[z_0,z_1,z_2,z_3,z_4]^\ZZZ =
  \bigoplus_{i=1}^{100}
  \eta_i 
  \, 
  \C[\theta_1,\theta_2,\theta_3,\theta_4,\theta_5 ]
  .
\end{equation}
Here, the primary invariants are~\cite{Braun:2007sn}
\begin{equation}
  \label{eq:thetadef}
  \begin{array}{rcl}
    \theta_1 \eqdef&\; 
    z_0^5+z_1^5+z_2^5+z_3^5+z_4^5 
    \;&= z_0^5 + \cyclperm
    \\
    \theta_2 \eqdef&\; 
    z_0 z_1 z_2 z_3 z_4 
    \\
    \theta_3 \eqdef&\; 
    z_0^3 z_1 z_4+z_0 z_1^3 z_2+z_0 z_3 z_4^3
    +z_1 z_2^3 z_3+z_2 z_3^3 z_4 
    \;&= z_0^3 z_1 z_4 + \cyclperm
    \\
    \theta_4 \eqdef&\; 
    z_0^{10}+z_1^{10}+z_2^{10}+z_3^{10}+z_4^{10} 
    \;&= z_0^{10} + \cyclperm
    \\
    \theta_5 \eqdef&\; 
    z_0^8 z_2 z_3+z_0 z_1 z_3^8+z_0 z_2^8 z_4+z_1^8 z_3 z_4+z_1 z_2
    z_4^8
    \;&= z_0^8 z_2 z_3 + \cyclperm
  \end{array}
\end{equation}
and the secondary invariants are, in degrees $< 10$, 
\begin{equation}
\begin{gathered}
  \begin{aligned}
    \eta_1 \eqdef&\; 1    ,\\ 
  \end{aligned}
  \\[1ex]
  \begin{aligned}
    \eta_2 \eqdef&\; z_0^2 z_1 z_2^2 + \cyclperm    ,&
    \eta_3 \eqdef&\; z_0^2 z_1^2 z_3 + \cyclperm    ,&
    \eta_4 \eqdef&\; z_0^3 z_2 z_3 + \cyclperm      
    .
  \end{aligned}
\end{gathered}
\end{equation}
Hence, an invariant quintic is of the form
\begin{equation}
  Q(z) = c_0 \theta_1 + c_1 \theta_2 + c_2 \theta_3 + 
  c_3 \eta_2 + c_4 \eta_3 + c_5 \eta_4
  .
\end{equation}
The complex structure moduli space of the invariant quintics is
an open subset of $\CP^5$ parametrized by $[c_0:\cdots:c_5]$. Finally,
one can easily check that a generic member of this family of quintics
is smooth and fixed-point free. Therefore, the quotient
\begin{equation}
  X = Q \Big/ \big(\ZZZ\big)
\end{equation}
is a smooth Calabi-Yau threefold with fundamental group $\pi_1(X) =
\ZZZ$.

\section{Monadology}
\label{sec:monad}

\subsection{Overview}
\label{sec:monadintro}

A monad~\cite{Monadology, Witten:1993yc, Distler:1993mk, Blumenhagen:1996vu} is a
three-step filtration of a vector bundle $V$. That is, a complex
\begin{equation}
  \xymatrix{
    0 \ar[r] & 
    A \ar[r]^-a & 
    B \ar[r]^-b & 
    C \ar[r] & 
    0
  }
\end{equation}
whose cohomology is a vector bundle in the middle, at the $B$
entry. In other words, $a$ is injective, $b$ is surjective, $b \circ
a = 0$, and $V = \ker(b)/\img(a)$ is the vector bundle. For example, a
short exact sequence corresponds to the zero vector bundle.

For the purposes of this paper, I will only consider \emph{positive
  monads} where $A=0$ and $B$, $C$ are very ample bundles. In this
case, $V$ \emph{is} defined by a short exact sequence
\begin{equation}
  \label{eq:posmonad1}
  \xymatrix{
    0 \ar[r] & 
    V \ar[r] & 
    B \ar[r]^-f & 
    C \ar[r] & 
    0
  }  
\end{equation}
In particular, I will always take the base space to be $Q\subset
\CP^4$ and 
\begin{equation}
  \label{eq:posmonad2}
  B = \bigoplus_{i=1}^n \Osheaf(b_i)
  ,\quad
  C = \bigoplus_{j=1}^m \Osheaf(c_j)
\end{equation}
with $b_i$, $c_j>0$. The positive monads for rank $3$, $4$, and $5$
bundles satisfying heterotic anomaly cancellation without anti-branes
were classified in~\cite{Anderson:2007nc, Anderson:2008uw,
  Anderson:2008ex}. There are $43$ such monads, none of which gives
rise to $3$ generations\footnote{The net number of generations, that
  is the difference between generations and anti-generations, equals
  the index $\Ind(V)$ of the vector bundle.}. Adding the additional
constraint that the number of generations is at least a multiple of
$3$, the authors
\begin{table}[htb]
  \begin{center}
    \begin{tabular}{|c|c|c|c|c|} \hline
      $\rank V$ & $\{b_i\}$ & $\{c_i\}$ & $\Ind V$ 
      & $G_\text{3-gen}$ \\ \hline\hline
      3 & 
      \textcolor{red}{$\{$2, 2, 1, 1, 1$\}$} & 
      \textcolor{red}{$\{$4, 3$\}$} & 
      \textcolor{red}{$-60$} & -- \\
      3 &
      \textcolor{red}{$\{$2, 2, 2, 1, 1$\}$} &
      \textcolor{red}{$\{$5, 3$\}$} & 
      \textcolor{red}{$-105$} & -- \\ 
      3 & 
      \textcolor{red}{$\{$3, 2, 1, 1, 1$\}$} & 
      \textcolor{red}{$\{$4, 4$\}$} & 
      $-75$ & -- \\ 
      3 & 
      % FIXME: Mention that this is Z_5 equivariant three-gen
      \textcolor{red}{$\{$1, 1, 1, 1, 1, 1$\}$} & 
      \textcolor{red}{$\{$2, 2, 2$\}$} & 
      \textcolor{red}{$-15$} & $\Z_5$ \\ 
      3 & 
      \textcolor{red}{$\{$2, 2, 2, 1, 1, 1$\}$} & 
      \textcolor{red}{$\{$3, 3, 3$\}$} & 
      \textcolor{red}{$-45$} & -- \\ 
      3 & 
      \textcolor{red}{$\{$3, 3, 3, 1, 1, 1$\}$} & 
      \textcolor{red}{$\{$4, 4, 4$\}$} & 
      \textcolor{red}{$-90$} & -- \\ 
      3 & 
      \textcolor{red}{$\{$2, 2, 2, 2, 2, 2, 2, 2$\}$} & 
      \textcolor{red}{$\{$4, 3, 3, 3, 3$\}$} &
      \textcolor{red}{$-90$} & -- \\ 
      3 & 
      \textcolor{red}{$\{$2, 2, 2, 2, 2, 2, 2, 2, 2$\}$} & 
      \textcolor{red}{$\{$3, 3, 3, 3, 3, 3$\}$} & 
      $-75$ & -- \\ 
      4 & 
      \textcolor{red}{$\{$2, 2, 1, 1, 1, 1$\}$} & 
      \textcolor{red}{$\{$4, 4$\}$} & 
      \textcolor{red}{$-90$} & -- \\ 
      4 & 
      \textcolor{red}{$\{$1, 1, 1, 1, 1, 1, 1$\}$} & 
      \textcolor{red}{$\{$3, 2, 2$\}$} & 
      \textcolor{red}{$-30$} & -- \\ 
      4 & 
      \textcolor{red}{$\{$2, 2, 2, 1, 1, 1, 1$\}$} & 
      \textcolor{red}{$\{$4, 3, 3$\}$} & 
      $-75$ & -- \\ 
      4 & 
      \textcolor{red}{$\{$2, 2, 2, 2, 1, 1, 1, 1$\}$} & 
      \textcolor{red}{$\{$3, 3, 3, 3$\}$} & 
      \textcolor{red}{$-60$} & -- \\ 
      5 & 
      \textcolor{red}{$\{$1, 1, 1, 1, 1, 1, 1, 1$\}$} & 
      \textcolor{red}{$\{$3, 3, 2$\}$} & 
      \textcolor{red}{$-45$} & -- \\ 
      5 & 
      \textcolor{red}{$\{$1, 1, 1, 1, 1, 1, 1, 1$\}$} & 
      \textcolor{red}{$\{$4, 2, 2$\}$} & 
      \textcolor{red}{$-60$} & -- \\ 
      5 & $\{$2, 2, 2, 2, 2, 1, 1, 1, 1, 1$\}$ & $\{$3, 3, 3, 3, 3$\}$ &
      $-75$ & $\Z_5\times\Z_5$\\ \hline
    \end{tabular}
  \end{center}
  \caption{Positive monad bundles on the quintic. The entries marked
    in \textcolor{red}{red} are obstructions to a $\ZZZ$-action, see
    \autoref{sec:topology}. The last column $G_\text{3-gen}$ are
    free group actions such that the quotient has three generations.}
  \label{tab:monads}
\end{table}
found $15$ monads on the quintic, which are reproduced in
\autoref{tab:monads} for convenience. As I will discuss shortly, most
of these monad bundles do not have a suitable symmetry for the
quotient to yield a three generation model. However, two of them
do. The first one is a rank $3$ bundle which we can divide by
$\Z_5$. This yields a low-energy $E_6$ gauge group with three
generations of $\Rep{27}$ and no anti-generations. The $\Z_5$
fundamental group can be used to break the $E_6$ to the standard model
gauge group plus two extra $U(1)$~\cite{Goodsell:2009xc}, though it
would be difficult to remove the exotic matter coming from the
decomposition of the $\Rep{27}$. The other model, and subject of this
paper, is in the last row of \autoref{tab:monads}. As we will see in
the remainder of this section, being able to divide out the non-cyclic
group $\ZZZ$ will pose many more restrictions on the monad bundle than
just $\Z_5$. At the same time, the larger group makes it much easier
to project out unwanted matter states.

\subsection{Symmetry Considerations}
\label{sec:symmetry}

For the reasons just mentioned, in the following I will only consider
the monad bundle\footnote{I will always write $r \Osheaf(n)$ for the
  rank-$r$ bundle $\oplus_{i=1}^r \Osheaf(n) = \Osheaf(n)^{\oplus r} =
  \Rep{r} \otimes \Osheaf(n)$.}
\begin{equation}
  \label{eq:monad}
  \xymatrix{
    0 \ar[r] & 
    V \ar[r] & 
    5 \Osheaf_Q(1) \oplus 5 \Osheaf_Q(2) \ar[r]^-f & 
    5 \Osheaf_Q(3) \ar[r] & 
    0
    .
  }  
\end{equation}
However, eq.~\eqref{eq:monad} only defines a vector bundle on
$Q\subset \CP^4$. In order to divide out the group action and obtain a
bundle on $X$, we need to first \emph{specify} a group action on $V$.

The choice of a group action on the vector bundle is called an
equivariant structure. This is a choice bundle map $\gamma(g):V\to V$
covering the group action $g:Q\to Q$ on the base of the bundle for all
group elements $g\in G$. In other words, we need to pick a linear map
$\gamma_p(g):V_p\to V_{gp}$ holomorphically varying over each point
$p\in Q$. Similarly, an equivariant map $f:(V,\gamma^V)\to
(W,\gamma^W)$ between equivariant bundles is a map of the vector
bundles that intertwines the equivariant structure in the obvious way,
\begin{equation}
  \label{eq:intertwine}
  \vcenter{\xymatrix{
      V_p \ar[r]^-{f_p(g)} \ar[d]_-{\gamma^V_p(g)} & 
      W_{p} \ar[d]^-{\gamma^W_p(g)} \\
      V_{gp} \ar[r]^-{f_{gp}(g)} & 
      W_{gp}
    }}
  \quad\Leftrightarrow\quad
  \gamma^W_p(g) \circ f_p(g) = f_{gp}(g) \circ \gamma^V_p(g)
  .
\end{equation}
Direct sums and tensor products of equivariant bundles are equivariant
in the obvious way, turning the category $\Vect_G$ of $G$-equivariant
vector bundles into a ring. Finally, a monad is equivariant if the
objects in the complex and the maps are equivariant; The cohomology of
an equivariant monad is an equivariant vector bundle.

For example, consider the trivial line bundle $\Osheaf$ on a compact
complex manifold. Up to an overall factor, there is a unique section
$s$ which is nowhere vanishing. Hence, the choice of a $G$-equivariant
structure on $\Osheaf$ is equivalent to the choice of a
$G$-representation on $\Gamma(\Osheaf)=\C s$, that is, a
multiplicative character of $G$. By abuse of
notation~\cite{Braun:2005zv}, we denote by $\chi \in
\Hom(G,\C^\times)$ also the corresponding line bundle. Moreover, we
write $\chi V$ for the tensor product of the line bundle $\chi$ and
the vector bundle $V$.

Since we are particularly interested in slope-stable bundles, let us
note that the equivariant structure on such a bundle is unique up to
multiplication with a multiplicative character. The proof is as
follows, assume that you have two different equivariant structures
$V_1=(V,\gamma_{(1)})$ and $V_2=(V,\gamma_{(2)})$. Then
$\gamma_{(2)}^{-1}\circ \gamma_{(1)}$ is a nontrivial automorphism of
$V$. But the only automorphisms of stable bundles are multiplication
by a constant~\cite{MR1600388}. The constant can depend on $g\in G$,
but must be a $1$-dimensional representation $\chi$. Therefore, $V_1 =
\chi V_2$.

\subsection{Schur Covers}
\label{sec:Schur}

Not every vector bundle can carry an equivariant structure. The
simplest example would be the line bundle $\Osheaf_Q(1)$. Let us look
at the problem in some detail. First, let us identify the sections
with the homogeneous coordinates on $\CP^4$ in the usual way,
\begin{equation}
  \Gamma \Osheaf(1) = 
  \langle z_0, z_1, z_2, z_3, z_4 \rangle
  .
\end{equation}
The naive guess for an equivariant structure would be the tautological
action
\begin{equation}
  \label{eq:tautological}
  \gamma(g_1)(z) = g_1(z)
  ,\quad
  \gamma(g_2)(z) = g_2(z)  
  .
\end{equation}
This fails to be a $\ZZZ$-equivariant structure because the two
actions do not commute,
\begin{equation}
  \gamma(g_2) \gamma(g_1) \gamma(g_2)^{-1} \gamma(g_1)^{-1} 
  = \zeta \not= 1
  ,
\end{equation}
see eq.~\eqref{eq:g1g2commutator}. We can define our way out of this
problem by introducing another generator $g_3$ such that
$\gamma(g_3)(z)=\zeta z$ and $g_2 g_1 g_2^{-1} g_1^{-1} = g_3$. This
enlarged group is the Heisenberg group
\begin{equation}
  H_5 = \langle g_1, g_2, g_3 \rangle = \big( \ZZZ \big) \rtimes \Z_5
  ,
\end{equation}
and we just defined a $H_5$-equivariant structure on
$\Osheaf_Q(1)$. We note that the Heisenberg group is a Schur cover of
$\ZZZ$,
\begin{equation}
  \label{eq:SchurCover}
  \xymatrix{
    0 \ar[r] &
    \Z_5 \ar[r] &
    H_5 \ar[r] &
    \ZZZ \ar[r] &
    0
  }
  ,
\end{equation}
and that this construction can be generalized: Given a projective
action on the homogeneous coordinates of a projective space,
$\Osheaf(1)$ is equivariant with respect to a Schur cover.

But we wanted a $\ZZZ$-equivariant structure, and not a
$H_5$-equivariant structure! Note that a $H_5$-equivariant structure
$\gamma$ is, in fact, a $\ZZZ$-equivariant structure if and only if
the kernel $\langle g_3 \rangle = \Z_5$ of the cover is represented
trivially. Therefore, we arrive at the following characterization of
$\ZZZ$-equivariant bundles: 
\begin{itemize}
\item Every vector bundle $(V,\gamma)$ is $H_5$-equivariant.
\item The vector bundle $(V,\gamma)$ is $\ZZZ$-equivariant if and only
  if\footnote{Note that $g_3$ acts trivially on the base space, so its
    action on the bundle is just a linear map of the fiber to itself.}
  $\gamma(g_3)=\Id$.
\end{itemize}
This shows that eq.~\eqref{eq:tautological} does not define an
$\ZZZ$-equivariant structure on $\Osheaf_Q(1)$. Could there be another
$H_5$-equivariant structure that \emph{does} descend to a
$\ZZZ$-equivariant structure? To rule out this possibility first note
that, like all line bundles, $\Osheaf_Q(1)$ is a slope-stable vector
bundle. By the argument in \autoref{sec:symmetry}, any other
equivariant structure must differ from~\eqref{eq:tautological} by a
multiplicative character of $H_5$. As we will see in more detail in
\autoref{sec:rep}, the multiplicative characters of $H_5$ are just the
$1$-d representations of $\ZZZ$. Therefore, any character $\chi$ of
$H_5$ satisfies $\chi(g_3)=1$ and we clearly cannot use this freedom
to turn eq.~\eqref{eq:tautological} into a $\ZZZ$-equivariant
structure. Hence, there cannot be any $\ZZZ$-equivariant structure on
$\Osheaf_Q(1)$.

\subsection{Topological Restrictions}
\label{sec:topology}

Although we have used the holomorphic structure in the above argument,
one can exclude the existence of a $\ZZZ$-equivariant structure on
$\Osheaf_Q(1)$ on grounds of topology alone. Recall that the
topological isomorphism class of a line bundle is classified by its
first Chern class. In the case at hand, it is
\begin{equation}
  c_1\big( \Osheaf_Q(1) \big) = 1 \in \Z = H^2\big(Q,\Z\big)
  .
\end{equation}
The cohomology of the cover $Q$ and the quotient $X=Q/(\ZZZ)$ is
related via the Leray-Serre spectral sequence,
\begin{equation}
  E_2^{p,q} = H^q\Big(Q, H^p(\ZZZ,Z) \Big)
  \quad \Rightarrow \quad
  H^{p+q}_{\ZZZ}(Q,\Z) = H^{p+q}(X,\Z)
  .
\end{equation}
For our purposes, the important part is the
nonvanishing~\cite{Aspinwall:1994uj, Braun:2007tp, Braun:2007xh,
  Braun:2007vy} differential $d_3$ in the tableau
\begin{equation}
  \label{eq:GLSssE2}
  E_2^{p,q} =
  \vcenter{
    \xymatrix@=2mm{
      {\scriptstyle \vdots}\hspace{1.6mm} &
      \vdots & \vdots & \vdots &\vdots & \vdots & \vdots & \adots \\
      {\scriptstyle q=2}\hspace{1.6mm} &
      \Z \ar[ddrrr]|(0.51){d_3} & 0 & \Z_5^2 & \Z_5 & 
      \Z_5^3 & \Z_5^2
      & \cdots \\
      {\scriptstyle q=1}\hspace{1.6mm} &
      0 & 0 & 0 & 0 & 0 & 0
      & \cdots \\
      {\scriptstyle q=0}\hspace{1.6mm} &
      \Z & 0 & \Z_5^2 & \Z_5 & 
      \Z_5^3 & \Z_5^2
      & \cdots \\
      \ar[]+/r 3.4mm/+/u 1.7mm/;[rrrrrrr]+/r 3mm/+/r 3.4mm/+/u 1.7mm/
      \ar[]+/r 3.4mm/+/u 1.7mm/;[uuuu]+/u  2mm/+/r 3.4mm/+/u 1.7mm/
      & 
      {\vbox{\vspace{3.5mm}}\scriptstyle p=0} & 
      {\vbox{\vspace{3.5mm}}\scriptstyle p=1} & 
      {\vbox{\vspace{3.5mm}}\scriptstyle p=2} & 
      {\vbox{\vspace{3.5mm}}\scriptstyle p=3} & 
      {\vbox{\vspace{3.5mm}}\scriptstyle p=4} & 
      {\vbox{\vspace{3.5mm}}\scriptstyle p=5} & 
      {\vbox{\vspace{3.5mm}}\scriptstyle \cdots }&
    }}
  .
\end{equation}
Therefore, only the multiples of $5 \in H^2(Q,\Z)$ survive to the
$E_3$ tableau\footnote{And, since the Leray-Serre spectral sequence is
  a first quadrant spectral sequence, $E_3^{0,2}=E_\infty^{0,2}$.}. In
particular, we obtain
\begin{equation}
  H^2\big(X,\Z)
  = 
  \ker(d_3) \oplus \Z_5 \oplus \Z_5
  ~\simeq~
  \Z \oplus \Z_5 \oplus \Z_5
\end{equation}
on the quotient\footnote{That is, the line bundles on the quotient are
  classified by the de Rham part of the first Chern class
  $\tfrac{1}{2\pi}[F]$ together with two $\Z_5$ phases for the
  discrete Wilson lines.}. Another way of expressing this factor of
$5$ is the following. Consider the quotient map $q:Q\to X$, and pull
back the cohomology groups. Then
\begin{equation}
  \label{eq:q2}
  q^\ast: 
  H^2\big(X,\Z)
  \longrightarrow
  H^2\big(Q,\Z)
  ,\quad
  (f,t_1,t_2)\mapsto 5 f
\end{equation}
is multiplication by $5$.

Now we apply the usual tautology 
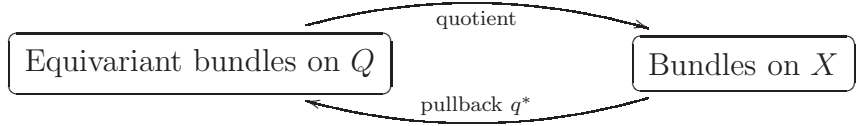
\begin{displaymath}
  \xymatrix@C=3cm{
    *++[F-:<3pt>]+{\text{Equivariant bundles on $Q$}}
    \ar@/^8mm/[r]_-{\text{quotient}}
    &
    *++[F-:<3pt>]+{\text{Bundles on $X$}}
    \ar@/^8mm/[l]_-{\text{pullback $q^*$}}
  }
\end{displaymath}
that equivariant bundles on $Q$ are the same as bundles on the
quotient $X$. The Chern classes are natural, that is, any bundle $W$
on $X$ satisfies $c_i \big( q^*W \big) = q^* \big( c_i(W)
\big)$. Therefore, a necessary condition for a bundle $V$ on $Q$ to
have a $\ZZZ$-equivariant structure is that its Chern classes are in the
image of $q$, that is,
\begin{equation}
  V\in \Vect_\ZZZ(Q) 
  \quad\Rightarrow\quad
  c_i(V) \in \img(q^*)
  .
\end{equation}
In particular, $c_1(V) \in 5 \Z \subset H^2(Q,\Z)$ is a necessary
condition for a bundle on $Q$ to be $\ZZZ$-equivariant. This
restriction on the first Chern class already forbids equivariant
structures on most of the $B$, $C$ bundles in \autoref{tab:monads}.

In fact, next to the monad bundle that I am investigating in this
paper, only the monad with $B=3 \Osheaf(2) \oplus 4 \Osheaf(1)$ and
$C=\Osheaf(4)\oplus 2\Osheaf(3)$ seems to be allowed. However, the
latter is excluded by its second Chern classes. For completeness, the
expressions for the integrally normalized Chern classes and the
pullback $q^*: H^\text{ev}(X)\to H^\text{ev}(Q)$ are in
\autoref{sec:Chern}.

\subsection{Heterotic Anomaly Cancellation}
\label{sec:anomaly}

Of course, the heterotic anomaly cancellation condition
\begin{equation}
  c_2(X) - c_2(W) - c_2(W_\text{hidden}) = PD(C) 
  \quad \in H^4\big(X,\Z\big)
\end{equation}
has to be satisfied on the quotient manifold $X=Q/(\Z_5\times\Z_5)$
and bundle $W=V/\gamma$, where $PD(C)$ is the Poincar\'e-dual of the
curve wrapped by five-branes. 

A necessary but not sufficient criterion for the anomaly cancellation
is that the image of both sides under the pull-back is the same,
\begin{equation}
  c_2(Q) - c_2(V) - c_2(V_\text{hidden}) = q^* PD(C)   
  \quad \in H^4\big(Q,\Z\big).
\end{equation}
However, on the quintic we are in the favorable circumstance that 
\begin{equation}
  q^*: 
  \underbrace{H^4(X,\Z)}_{\simeq \Z} 
  \to
  \underbrace{H^4(Q,\Z)}_{\simeq \Z}
  ,\quad
  n\mapsto 25 n
\end{equation}
is injective\footnote{Modulo torsion (the finite part in $H^4$), the
  pull-back is always injective. However, it is important to cancel it
  in integral cohomology on the quotient manifold as there is a danger
  of a discrete anomaly. Moreover, the existence of holomorphic curves
  can depend on the torsion part of their homology
  class~\cite{Braun:2007tp, Braun:2007xh, Braun:2007vy}.}. Hence, the
anomaly cancellation condition on the covering space and on the
quotient are equivalent. In particular, the monad bundle
eq.~\eqref{eq:monad} satisfies $c_2(V)=c_2(Q)$, and therefore cancels
the heterotic anomaly on the cover $Q$ as well as on the quotient $X$
without hidden bundle or branes.

\subsection{Representation Theory}
\label{sec:rep}

In order to better understand the Schur cover
\begin{equation}
  H_5 = \big< g_1, g_2, g_3
  \big|
  g_1^5=g_2^5=g_3^5=1
  ,~
  g_2g_1g_2^{-1}g_1^{-1}=g_3
  ,~
  g_1g_3=g_3g_1
  ,~
  g_2g_3=g_3g_2
  \big>
\end{equation}
let us quickly discuss its representation theory~\cite{GAP4}. First of
all, its Abelianisation is $\ZZZ$. Therefore, each $1$-dimensional
representation has to factor through $\ZZZ$, see
eq.~\eqref{eq:SchurCover}, and $H_5$ has $25$ one-dimensional
representations $r_1^i r_2^j$, $0\geq i,j<5$. In addition, there are
four\footnote{In the following, we will use the notation where these
  four representations are indexed by $\Z\text{ mod }5$, with $0$
  being disallowed. In other words, $R_{i+5}\eqdef R_i$.} useful to
take irreducible representations $R_1$,~$\dots$,~$R_4$ of dimension
$5$, distinguished by the weight
\begin{equation}
  R_i(g_3) = \zeta^i \diag(1,1,1,1,1)
  .
\end{equation}
For example, we defined the representation carried by the $5$
homogeneous variables of $\CP^4$ to be $R_1$, see
eq.~\eqref{eq:gzact}. Together, these are all irreducible
representations:
\begin{equation}
  \sum_{i,j=0}^4 \dim\big(r_1^i r_2^j\big)^2 +
  \sum_{i=1}^4 \dim\big(R_i\big)^2 
  = 125 = |H_5|
  .
\end{equation}
The tensor products of the irreducible representations can be
summarized in the representation ring
\begin{multline}
  R(H_5) = \Z\big[r_1,r_2,R_1,R_2,R_3,R_4\big]
  \Big/
  \Big< 
  r_i^5=1
  ,\quad
  r_i R_j=R_j
  ,~
  \\
  R_a R_{5-a} = \sum r_1^ir_2^j
  ,\quad
  R_a R_b = 5 R_{a+b}\text{ if $a+b\not=0$ mod $5$}
  \Big>
  .
\end{multline}
One observes that
\begin{itemize}
\item Take one of the genuinely $H_5$ representations $R_i$.
\item Tensor with $R_{5-i}$.
\item The result is a $\ZZZ$ representation, that is, $R_i
  R_{5-i}(g_3)=\Id$.
\end{itemize}
This will be important in the following to construct
$\ZZZ$-equivariant rank-$5$ vector bundles. 

Finally, since we will be interested in polynomials, we will need the
symmetric powers of $R_1$. They are
\begin{equation}
  \label{eq:SymR1}
  \Sym^k(R_1) = 
  \begin{cases}
    \frac{1}{5}\binom{4+k}{k} R_k
    &
    k\not=0\text{ mod }5
    ,
    \\
    1 + 
    \frac{1}{25}\Big[\binom{4+k}{k}-1\Big]
    \sum_{i,j} r_1^i r_2^j
    &
    k=0\text{ mod }5
    .
  \end{cases}
\end{equation}

\subsection{Equivariance}
\label{sec:equivariant}

Every representation $\rho \in R(H_5)$ of dimension $r=\dim(\rho)$
defines an $H_5$-equivariant vector bundle on $Q$ by taking the
trivial vector bundle $r \Osheaf_Q$ with its global sections
$\vec{s}=(s_\alpha)$, $\alpha\in\{0,\dots,r-1\}$, and defining an
equivariant structure
\begin{equation}
  \gamma(g)(\vec{s}) = \rho(g) \vec{s}
  .
\end{equation}
By abuse of notation, I will denote the corresponding vector bundle by
$\rho$ as well. In particular, taking $\rho=R_{5-n}$ and tensoring
with a line bundle defines the rank $5$ vector bundles
\begin{equation}
  \label{eq:Phi}
  \Phi(n) \eqdef R_{5-n} \otimes \Osheaf_Q(n)
  ,\qquad
  n\not= 0\text{ mod }5
  .
\end{equation}
Previously, in \autoref{sec:Schur}, I defined a $H_5$-equivariant
structure on $\Osheaf_Q(1)$ and, hence, on
$\Osheaf_Q(n)=\Osheaf_Q(1)^{\otimes n}$. Therefore, $\Phi(n)$ is
$H_5$-equivariant as the tensor product of equivariant
bundles. Moreover, $g_3$ acts trivially, and $\Phi(n)$ is actually a
$\ZZZ$-equivariant vector bundle. Topologically $\Phi(n)=5
\Osheaf_Q(n)$ is decomposable, but as an equivariant bundle it is not.

Let us take a closer look at this definition. A basis for
$H^0\big(Q,\Phi(n)\big)$ is the same as for $5\Osheaf_Q(n)$, namely
the $5\cdot \binom{4+n}{n}$ sections
\begin{equation}
  s_{\alpha,(i_1,\dots,i_n)}
  \eqdef
  \begin{pmatrix}
    0 \\ \vdots \\ 0 \\ 
    z_{i_1} z_{i_2} \cdots z_{i_n} 
    \mathrlap{\textcolor{blue}{
        \qquad\longleftarrow\text{row $\alpha$}
      }}
    \\ 0 \\ \vdots \\ 
    0\mathrlap{\hspace{13mm}, \qquad  0\leq \alpha, i_1,\dots, i_n<5.}
  \end{pmatrix}
  \phantom{, \qquad  0\leq \alpha, i_1,\dots, i_n<5.}
\end{equation}
The generators $g_1$, $g_2$ act in an obvious way on the homogeneous
coordinates $z_0$, $\dots$, $z_4$, this is the $H_5$ action on
$\Osheaf_Q(1)$. The action on the sections of $\Phi(n)$ is slightly
different, and a convenient basis choice is
\begin{equation} 
  \label{eq:g1g2s}
  \begin{split}
    g_1\big( s_{\alpha,(i_1,~\dots,~i_n)} \big)
    \eqdef&\;
    s_{\alpha-n, (i_1+1,~\dots,~ i_n+1)}
    ,
    \\
    g_2\big( s_{\alpha,(i_1,~\dots,~i_n)} \big)
    \eqdef&\;
    \zeta^{\alpha+i_1+\cdots+i_n}
    s_{\alpha, (i_1,~\dots,~ i_n)}    
    .
  \end{split}
\end{equation}
One can easily check that
\begin{equation}
  g_1\circ g_2 
  \big( s_{\alpha,(i_1,~\dots,~i_n)} \big) 
  = 
  \zeta^{\alpha+i_1+\cdots+i_n}
  s_{\alpha-n, (i_1+1,~\dots,~ i_n+1)}
  =
  g_2\circ g_1
  \big( s_{\alpha,(i_1,~\dots,~i_n)} \big)   
  .
\end{equation}
Therefore, eq.~\eqref{eq:g1g2s} defines a $\ZZZ$-equivariant structure
on $\Phi(n)$. 

To summarize, we have now defined equivariant structures on the
entries of the monad under consideration. They are
\begin{equation}
  B = \Phi(1) \oplus \Phi(2)
  ,\quad
  C = \Phi(3)
  .
\end{equation}
If we can find a equivariant morphism from $B$ to $C$, then we have
defined an equivariant monad.

\subsection{Morphisms}
\label{sec:map}

A morphism $B\to C$ is simply given by a $\rank(C)\times \rank(B)$
matrix of polynomials such that the $(j,i)$ entry is of degree
$c_j-b_i$. Keeping track of the equivariant structure on the bundles
$\Phi(n)$ and assuming\footnote{I will allways assume this in the
  following. For the purposes of this paper, only the case where
  $1\geq n,m\geq 3$ is relevant.} $m-n\text{ mod }5\not=0$, one finds
\begin{multline}
  \Hom\big(\Phi(n),\Phi(m) \big) =
  \Hom\Big(R_{5-n}\Osheaf_Q(n),R_{5-m}\Osheaf_Q(m) \Big) 
  \\
  =
  R_{5-n}^\dual R_{5-m} \Sym^{m-n}(R_1) 
  = 
  \binom{4+m-n}{m-n} \sum_{i,j=0}^4 r_1^i r_2^j
  .
\end{multline}

Let me describe the invariant homomorphisms in more detail. First,
note the obvious basis
% \begin{equation}
%   \Hom\big( \Phi(n),\Phi(n+k) \big) =
%   \big<
%   f^\alpha_{\beta,(i_1,\dots,i_k)}
%   \big>
%   ,\qquad
%   f^\alpha_{\beta,(i_1,\dots,i_k)} \eqdef
%   (\delta^\alpha_\beta) \;
%   z_{i_1}\cdots z_{i_k}
% \end{equation}
\begin{multline}
  \Hom\big( \Phi(n),\Phi(n+k) \big) =
  \big<
  f^\alpha_{\beta,(i_1,\dots,i_k)}
  \big>
  ,
  \\[3ex]
  f^\alpha_{\beta,(i_1,\dots,i_k)} \eqdef
  (\delta^\alpha_\beta) \;
  z_{i_1}\cdots z_{i_k}
  =
  \begin{pmatrix}
    0      & \cdots & 
    \mathclap{\textcolor{blue}{\smash{
          \begin{array}[b]{ccc}
            \text{column $\alpha$}\\
            \downarrow\\
            \\
          \end{array}
        }}}
    \mathclap{0}                  & \cdots & 0 \\
    \vdots &   &                    &        & \vdots \\
    0      &        & z_{i_1}\cdots z_{i_k} &        & 0
    \mathrlap{
      \textcolor{blue}{
        \quad \longleftarrow \text{ row $\beta$}
      }
      ,
    } \\
    \vdots &        &                    &  & \vdots \\
    0      & \cdots & 0                  & \cdots & 0 
  \end{pmatrix}
  \phantom{\quad \longleftarrow \text{ row $\alpha$}}
\end{multline}
where $(\delta^\alpha_\beta)$ is the $5\times 5$-matrix with a single
non-zero entry $=1$ and the $z_i$ are again the homogeneous
coordinates of $\CP^4$. The action on a section of $\Phi(n)$ is simply
matrix-vector multiplication, that is,
\begin{equation}
  f^\alpha_{\beta,(i_{n+1},\dots,i_{n+k})} 
  \big( s_{\epsilon,(i_1,\dots,i_n)} \big)
  = 
  \delta^\alpha_\epsilon 
  s_{\beta,(i_1,\dots,i_{n+k})}
  .
\end{equation}
Finally, the $g_1$, $g_2$ action on the space of maps is the usual
action on the homogeneous coordinates combined with a matrix action to
correctly intertwine between $\Phi(n)$ and $\Phi(n+k)$, see
eq.~\eqref{eq:intertwine}. Explicitly, the action is
\begin{equation}
  \begin{split}
    g_1\big( f^\alpha_{\beta,(i_1,\dots,i_k)}  \big)
    =&\; 
    f^{\alpha-n}_{\beta-n-k,(i_1+1,\dots,i_k+1)}  
    ,\\
    g_2\big( f^\alpha_{\beta,(i_1,\dots,i_k)}  \big)
    =&\; 
    \zeta^{\beta-\alpha+i_1+\cdots+i_k}
    f^\alpha_{\beta,(i_1,\dots,i_k)}  
    .
  \end{split}
\end{equation}
The $g_1$ and $g_2$ actions commute,
\begin{equation}
  g_2 g_1\big( f^\alpha_{\beta,(i_1,\dots,i_k)}  \big)
  =
  \zeta^{\beta-\alpha+i_1+\cdots+i_k}
  f^{\alpha-n}_{\beta-n-k,(i_1+1,\dots,i_k+1)}  
  =
  g_1 g_2 \big( f^\alpha_{\beta,(i_1,\dots,i_k)}  \big)
  ,
\end{equation}
and therefore decompose into $\ZZZ$-representations as we argued
above. For example, the $5$-dimensional space of invariant
homomorphisms from $\Phi(2)$ to $\Phi(3)$ is spanned by
% from $\Phi(1)$ to $\Phi(2)$
% \begin{equation}
%   \left(\begin{smallmatrix}
%       z_0 &   0 &   0 &   0 &   0 \\
%       0   &   0 &   0 & z_2 &   0 \\
%       0   & z_4 &   0 &   0 &   0 \\
%       0   &   0 &   0 &   0 & z_1 \\
%       0   &   0 & z_3 &   0 &   0 \\
%     \end{smallmatrix}\right)
%   ,
%   \left(\begin{smallmatrix}
%       0   & z_1 &   0 &   0 &   0 \\
%       0   &   0 &   0 &   0 & z_3 \\
%       0   &   0 & z_0 &   0 &   0 \\
%       z_2 &   0 &   0 &   0 &   0 \\
%       0   &   0 &   0 & z_4 &   0 \\
%     \end{smallmatrix}\right)
%   ,
%   \left(\begin{smallmatrix}
%       0   & 0   & z_2 &   0 &   0 \\
%       z_4 & 0   &   0 &   0 &   0 \\
%       0   & 0   &   0 & z_1 &   0 \\
%       0   & z_3 &   0 &   0 &   0 \\
%       0   & 0   &   0 &   0 & z_0 \\
%     \end{smallmatrix}\right)
%   ,
%   \left(\begin{smallmatrix}
%       0   & 0   & 0   & z_3 &   0 \\
%       0   & z_0 & 0   &   0 &   0 \\
%       0   & 0   & 0   &   0 & z_2 \\
%       0   & 0   & z_4 &   0 &   0 \\
%       z_1 & 0   & 0   &   0 &   0 \\
%     \end{smallmatrix}\right)
%   ,
%   \left(\begin{smallmatrix}
%       0   & 0   & 0   & 0   & z_4 \\
%       0   & 0   & z_1 & 0   &   0 \\
%       z_3 & 0   & 0   & 0   &   0 \\
%       0   & 0   & 0   & z_0 &   0 \\
%       0   & z_2 & 0   & 0   &   0 \\
%     \end{smallmatrix}\right)
%   .
% \end{equation}
\begin{equation}
  \left(\begin{smallmatrix}
      z_0& 0& 0& 0& 0 \\ 
      0& 0& 0& 0& z_3 \\ 
      0& 0& 0& z_1& 0 \\ 
      0& 0& z_4& 0& 0 \\ 
      0& z_2& 0& 0& 0
    \end{smallmatrix}\right)
  ,
  \left(\begin{smallmatrix}
      0& z_1& 0& 0& 0 \\ 
      z_4& 0& 0& 0& 0 \\ 
      0& 0& 0& 0& z_2 \\ 
      0& 0& 0& z_0& 0 \\ 
      0& 0& z_3& 0& 0
    \end{smallmatrix}\right)
  ,
  \left(\begin{smallmatrix}
      0& 0& z_2& 0& 0 \\ 
      0& z_0& 0& 0& 0 \\ 
      z_3& 0& 0& 0& 0 \\ 
      0& 0& 0& 0& z_1 \\ 
      0& 0& 0& z_4& 0
    \end{smallmatrix}\right)
  ,
  \left(\begin{smallmatrix}
      0& 0& 0& z_3& 0 \\ 
      0& 0& z_1& 0& 0 \\ 
      0& z_4& 0& 0& 0 \\
      z_2& 0& 0& 0& 0 \\ 
      0& 0& 0& 0& z_0
    \end{smallmatrix}\right)
  ,
  \left(\begin{smallmatrix}
      0& 0& 0& 0& z_4 \\
      0& 0& 0& z_2& 0 \\
      0& 0& z_0& 0& 0 \\ 
      0& z_3& 0& 0& 0 \\ 
      z_1& 0& 0& 0& 0  
    \end{smallmatrix}\right)
  .
\end{equation}
We now have all ingredients to define a $\ZZZ$-equivariant monad
bundle on the quintic. For explicitness, I will from now on take the
quintic to be the Fermat quintic
\begin{equation}
  Q = z_0^5+z_1^5+z_2^5+z_3^5+z_4^5
%  Q = u^5+v^5+x^5+y^5+z^5
  .
\end{equation}
Let the $\ZZZ$-equivariant rank-$5$ vector bundle be the kernel of the
positive monad
\begin{equation}
  \label{eq:ZZZmonad}
  \xymatrix{
    0 \ar[r] & 
    V \ar[r] & 
    \Phi(1) \oplus \Phi(2) \ar[r]^-f & 
    \Phi(3) \ar[r] & 
    0
  }  
\end{equation}
with an invariant map given by the polynomial matrix
\begin{equation}
  \label{eq:fdef}
  f \eqdef
  \begin{pmatrix}
    0 &  z_3^2 & 0 &  0 &  z_2^2 & 0 & z_1 & 0 & z_3 & z_4 \\
    0 &  z_0^2 & 0 &  z_1^2 & 0 &  z_4 & 0 & z_1 & z_2 & 0 \\
    z_4^2 & 0 &  0 &  z_3^2 & 0 &  0 & z_4 & z_0 & 0 & z_2 \\
    z_1^2 & 0 &  z_2^2 & 0 &  0 &  z_2 & z_3 & 0 & z_0 & 0 \\
    0 &  0 &  z_4^2 & 0 &  z_0^2 & z_1 & 0 & z_3 & 0 & z_0 
    % 0 &  y^2 & 0 &  0 &  x^2 & 0 & v & 0 & y & z \\
    % 0 &  u^2 & 0 &  v^2 & 0 &  z & 0 & v & x & 0 \\
    % z^2 & 0 &  0 &  y^2 & 0 &  0 & z & u & 0 & x \\
    % v^2 & 0 &  x^2 & 0 &  0 &  x & y & 0 & u & 0 \\
    % 0 &  0 &  z^2 & 0 &  u^2 & v & 0 & y & 0 & u 
  \end{pmatrix}
  .
\end{equation}
As we will see soon, the map $f$ has been chosen generic enough so
that the monad is, indeed, a vector bundle. Yet it is special enough
so that one $\Rep{5}$--$\barRep{5}$ pair survives\footnote{The
  observation of~\cite{Anderson:2007nc, Anderson:2008uw,
    Anderson:2008ex} that a completely generic map leads to no
  vector-like pairs whatsoever remains true.}.

% cat <<EOF | sed  -e's/z/z_4/g' -e's/u/z_0/g' -e's/v/z_1/g' -e's/x/z_2/g' -e's/y/z_3/g'

\section{Particle Spectrum}
\label{sec:spectrum}

\subsection{Machinery}
\label{sec:SS}

Massless fields in $10$ dimensions give rise to massless fields in the
$4$-dimensional effective action if their dependence on the Calabi-Yau
coordinates is a zero mode of the Dirac operator. This boils down to a
question about harmonic forms on the vector bundle $V$ and its
exterior powers $\wedge^i V$, $1\geq i\geq 4$. By changing our model
for the cohomology to sheaf cohomology, this becomes a tractable
computation. In particular, we can use the defining monad to convert
$\wedge^i V$ into something that depends only on $B$ and $C$ and the
map in-between. But since $B$ and $C$ are only sums of line bundles,
all cohomology groups are now given explicitly as vector spaces
spanned by polynomials.

Explicitly, we first replace $\wedge^i V$ by equivalent objects in the
derived category,\footnote{The underlined entry marks the zero
  position of the complex.}
\begin{equation}
  \label{eq:DerV}
  \begin{split}
    V =&\;
    \left[\strut
      0 \longrightarrow
      \underline{B} \stackrel{f}{\longrightarrow} C
      \longrightarrow
      0
    \right]
    ,
    \\
    \wedge^2 V =&\;
    \left[\strut
      0 \longrightarrow
      \underline{\wedge^2B} 
      \xrightarrow{b_1\wedge b_2 \mapsto [b_1 \otimes f(b_2)]}
      B\otimes C  \
      \xrightarrow{(b,c)\mapsto \{f(b)\otimes c\} }
      \Sym^2 C
      \longrightarrow
      0
    \right]
    % ,
    % \\
    % \wedge^3 V =&\;
    % \left[\strut
    %   \underline{\wedge^3B} \
    %   \xrightarrow{\wedge^2\Id \otimes f}
    %   \wedge^2 B\otimes C  \longrightarrow 
    %   B\otimes \Sym^2 C  \longrightarrow 
    %   \Sym^3 C
    % \right]
    .
  \end{split}
\end{equation}
The remaining exterior powers are just duals, namely
\begin{equation}
  % \wedge^2 V = \wedge^3V^\dual
  % ,\quad
  \wedge^3 V = \wedge^2V^\dual
  ,\quad
  \wedge^4 V = V^\dual
  ,\quad
  \wedge^5 V = \det V = \Osheaf_Q
  ,
\end{equation}
and their cohomology can be determined via Serre duality. As a
necessary evil we have to deal with cohomology for complexes, the
so-called hypercohomology. Note that there are two ways to ``take
cohomology'' here: There is the cohomology of a complex
$\Ksheaf^\bullet$, and the cohomology of the objects in the
complex. Here, we only need the case where the cohomology of the
complex is a vector bundle (or sheaf) $V$ located at a single
position,
\begin{equation}
  V = 
  \ker(\Ksheaf^0\to \Ksheaf^1) \Big/
  \img(\Ksheaf^{-1})
  ,\qquad
  \ker(\Ksheaf^p\to \Ksheaf^{p+1}) = \img(\Ksheaf^{p-1})
  \text{ if }p\not=0
  .
\end{equation}
In this case, the hypercohomology $H(\Ksheaf^\bullet)$ is simply the
cohomology of $V$. The other way of taking cohomologies gives rise to
the hypercohomology spectral sequence
\begin{equation}
  E_1^{p,q} = H^q \big( Q, \Ksheaf^p \big)
  \quad \Rightarrow \quad 
  H^{p+q}(V)
\end{equation}
Thanks to Kodaira vanishing, only the $q=0$ row will be
nonzero. Moreover, the first and only non-vanishing differential
$d_1:E_1^{p,0}\to E_1^{p+1,0}$ is just multiplication by $H(f)$
induced from $f:B\to C$ with suitable (anti-)symmetrization.

To summarize, using the defining map $f$, see eq.~\eqref{eq:fdef},
defines maps between spaces of polynomials
\begin{equation}
  \begin{gathered}
    H^0\big( Q, B \big)
    \xrightarrow{H(f)}
    H^0\big( Q, C \big)
    ,
    \\[1ex]
    H^0\big( Q, \wedge^2B \big)
    \stackrel{F_B}{\longrightarrow}
    H^0\big( Q, B\otimes C \big)
    \stackrel{F_C}{\longrightarrow}
    H^0\big( Q, \Sym^2 C \big)
    .
  \end{gathered}
\end{equation}
Here, $H(f)$ is tautologically the same matrix as in
eq.~\eqref{eq:fdef}. The polynomial matrices\footnote{Of course they
  satisfy $F_C F_B = 0$, as required for the $d_1$ differentials in
  the Hypercohomology spectral sequence.} $F_B$ and $F_C$ are of
dimension $50\times 45$ and $15\times 50$, respectively, and are
explicitly constructed by the script in \autoref{sec:singular}. Taking
the cohomology, one finds
\begin{equation}
  H^i\big(Q,V\big) = 
  \begin{cases}
    0   \\
    0   \\
    \coker H(f) \\
    \ker H(f)  
  \end{cases}
  , \quad
  H^i\big(Q,\wedge^2 V\big)  =
  \begin{cases}
    0  & i=3 \\
    \coker F_C  & i=2 \\ 
    \ker F_C \big/ \img F_B & i=1 \\
    \ker F_B & i=0 \\ 
  \end{cases}
  .
\end{equation}
Finally, using Serre duality,
\begin{equation}
  H^i\big(Q,\wedge^k V\big) = H^{3-i}\big(Q,\wedge^{5-k} V\big)
  ,
\end{equation}
we determined all cohomology groups of exterior powers of $V$. The
relevant computations with polynomials can be easily done with
\Singular~\cite{GPS05} and are recorded in
\autoref{sec:singular}.

\subsection{Slope Stability}
\label{sec:stab}

Since the quintic has only a one-dimensional $H^2(Q,\Z)$ we are in the
lucky case where we can apply Hoppe's
criterion~\cite{MR757476}. Specifically, we have
\begin{itemize}
\item The Fermat quintic $Q$ is a smooth manifold with
  $\dim\,H^2(Q,\Z)=1$.
\item The monad defines a \emph{vector bundle} $V$; This requires a
  short computation that $Q$ and the $5\times 5$ minors of $f$ do not
  vanish simultaneously.
\item The first Chern class of $V$ vanishes by construction.
\item Finally, $H^0(Q,\wedge^iV)$ must vanish for $1\geq i\geq 4$. The
  potential contributions are $\ker H(f)$ and $\ker F_B$, and a short
  computation shows that they indeed vanish, see again
  \autoref{sec:singular}. This can also be argued more generally using
  the Koszul resolution~\cite{Anderson:2008uw, Anderson:2008ex}.
\end{itemize}
Hoppe's criterion then guarantees that the bundle $V$ is slope-stable
and, therefore, admits a Hermitian Yang-Mills connection.

\subsection{Light Matter}
\label{sec:matter}

Using the well-known embedding of $SU(5)\times SU(5) \subset E_8$, the
matter spectrum of the $E_8\times E_8$ heterotic string compactified
on a slope-stable vector bundle over a Calabi-Yau manifold is
determined by the multiplicities\footnote{The last equation follows
  from $\Ind(V)=\Ind(\wedge^2V)$ for a rank-$5$ bundle or anomaly
  cancellation in the low-energy action.}
\begin{equation}
  \begin{gathered}
    n_\Rep{10} = h^1 (V)
    ,\quad
    n_\barRep{10} = h^1 (V^\dual) = h^2(V)
    ,\quad
    \\[1ex]
    n_\Rep{5} = h^1 (\wedge^2 V^\dual) = h^2 (\wedge^2 V)
    ,\quad
    n_\barRep{5} = h^1 (\wedge^2 V ) = n_\Rep{10} - n_\barRep{10} + n_\Rep{5}
    .
  \end{gathered}
\end{equation}
We already noted that $h^0(V)$ had to vanish for
slope-stability. Therefore, index theory determines the only
non-vanishing cohomology group $h^1(V)=75$. Moreover, the
$\ZZZ$-action is uniquely determined by the corresponding
character-valued index, and one obtains
\begin{equation}
  H^1\big( Q, V \big) = 3 \sum_{i,j=0}^4 r_1^i r_2^j
  .
\end{equation}
Next, consider $H^2(Q,\wedge^2V) = \coker F_C$. We can think of it as
symmetric tensors in $H^0(Q,C\otimes C)$ with the basis
\begin{equation}
  \Big\{
  \C\big[z_0,z_1,z_2,z_3,z_4\big]_6 \Big/ 
  \langle Q=0 \rangle
  \Big\}
  \otimes 
  \big\{ \vec{e}_{(\alpha,\beta)} \big|~ 0\geq \alpha\geq \beta \geq 4 \big\}
\end{equation}
The $\ZZZ$-group action can easily be identified as the one coming
from
\begin{equation}
  \Gamma \Phi(3) \times \Gamma \Phi(3)
  \longrightarrow
  H^0\big( Q, \Sym^2 C \big)
  ,\quad
  \big( s_{\alpha,(i_1,i_2,i_3)}, s_{\beta,(i_4,i_5,i_6)} \big)
  \mapsto
  e_{(\alpha,\beta)} \prod_{k=1}^6 z_{i_k}
  .
\end{equation}
When asked nicely, \Singular can compute a basis for the
cokernel, see \autoref{sec:singular}. One obtains
\begin{equation}
  \coker F_C = 
  \big<
  z_4^6       \vec{e}_{(1,4)},~
  z_0^3 z_2 z_4^2 \vec{e}_{(0,0)},~
  z_3^6       \vec{e}_{(1,1)},~
  z_1 z_3^5     \vec{e}_{(2,2)},~
  z_4^6       \vec{e}_{(3,3)},~
  z_2 z_4^5     \vec{e}_{(4,4)}
  \big>.
\end{equation}
In general, the $\ZZZ$-action will map the representatives to
different representatives of the same quotient space. These must be
projected back onto the chosen representatives to read off the group
action. Decomposing this $6$-dimensional $\ZZZ$-representation into
irreducible representations yields
\begin{equation}
  H^1\big( Q, \wedge^2 V \big) = 
  H^2\big( Q, \wedge^2 V \big) = 
  r_2^4 + 1+r_1+r_1^2+r_1^3+r_1^4
  .
\end{equation}

To summarize, we defined an equivariant bundle $(V,\gamma)$ on $Q$ and
computed the $\ZZZ$-action on its cohomology. Now, finally, we extract
the invariants to obtain the cohomology of the quotient bundle
$W = V/\gamma$ on the quotient space $X=Q/(\ZZZ)$,
\begin{equation}
  H^i\big(X,W\big) = 
  \begin{cases}
    0   \\
    0   \\
    3   \\
    0
  \end{cases}
  , \quad
  H^i\big(X,\wedge^2 W\big)  =
  \begin{cases}
    0  & i=3 \\
    1  & i=2 \\ 
    1+3  & i=1 \\
    0  & i=0 \\ 
  \end{cases}
  .
\end{equation}
Therefore, the ensuing $SU(5)$ GUT has a matter spectrum of three
$\Rep{10}+\barRep{5}$ and a single vector-like pair
$\Rep{5}$--$\barRep{5}$. Perturbing the bundle moduli $\phi$ by
changing the map eq.~\eqref{eq:fdef} removes this vector-like pair, so
there must be a $\mu$-term $\sim \phi \psi_\Rep{5}
\psi_\barRep{5}$ in the superpotential.

\appendix
% Hack that fixes reference name:
\makeatletter
\def\Hy@chapterstring{section}
\makeatother

\section{Chern classes on the Quintic and Quotient}
\label{sec:Chern}

The even-degree cohomology of the quintic is one-dimensional in
degrees $0$ to $6$, but the integral normalization of the cup product
still provides some interesting structure. If we denote the (positive)
generator of $H^2(Q,\Z)$ by $J$, then
\begin{equation}
  H^\text{ev}(Q,\Z) = \Z\big[ J, \tfrac{1}{5} J^2 \big]
  = 
  \Z ~\oplus~
  \Z \cdot J ~\oplus~ 
  \Z \cdot \left(\tfrac{1}{5} J^2\right) ~\oplus~
  \Z \cdot \left(\tfrac{1}{5} J^3\right) 
  .
\end{equation}
In other words, the square $J^2\in H^4(J,\Z)$ can be divided by $5$ in
the integral cohomology. Therefore, the integrally normalized Chern
classes of a positive monad, see eq.~\eqref{eq:posmonad1}, with
$c_1(V)=0$ are
\begin{equation}
  \begin{aligned}
    \rank(B) =&\; n
    ,
    % & 
    % \rank(C) =&\; m
    &
    \qquad
    \rank(V) =&\; n - m    
    ,
    \\
    c_1(B) =&\;
    \sum_{i=1}^n b_i
    ,
    % &
    % c_1(C) =&\;
    % \sum_{j=1}^m c_j
    &
    c_1(V) =&\;
    \sum_{i=1}^n b_i - \sum_{j=1}^m c_j
    \stackrel{!}{=} 0
    ,
    \\
    c_2(B) =&\;
    5 \sum_{i<j} b_i b_j
    ,
    % &
    % c_2(C) =&\;
    % 5 \sum_{i<j} c_i c_j
    &
    c_2(V) =&\;
    - \frac{5}{2} \left(\sum_{i=1}^n b_i^2 - \sum_{j=1}^m c_j^2\right)
    ,
    \\
    c_3(B) =&\;
    5 
    \sum_{i<j<k} b_i b_j b_k
    ,
    % &
    % c_3(C) =&\;
    % 5 
    % \sum_{i<j<k} c_i c_j c_k
    &
    c_3(V) =&\;
    \frac{5}{3} \left(\sum_{i=1}^n b_i^3 - \sum_{j=1}^m c_j^3\right)
    .
  \end{aligned}
\end{equation}
The pull-back by the $\ZZZ$ quotient map $q:Q\to X$ is~\cite{Braun:2007xh}
\begin{equation}
  \begin{aligned}
    q^*:&&
    \Z =&\; H^0(X,\Z) \to H^0(Q,\Z) = \Z
    ,&
    n &\mapsto n
    \\
    q^*:&&
    \Z\oplus\Z_5\oplus\Z_5 =&\; H^2(X,\Z) \to H^2(Q,\Z) = \Z
    ,&
    (n,\psi_1,\psi_2) &\mapsto 5 n
    \\
    q^*:&&
    \Z =&\; H^4(X,\Z) \to H^4(Q,\Z) = \Z
    ,&
    n &\mapsto 25 n
    \\
    q^*:&&
    \Z =&\; H^6(X,\Z) \to H^6(Q,\Z) = \Z
    ,&
    n &\mapsto 25 n
  \end{aligned}
\end{equation}
Therefore, a necessary condition for $B$, $C$ to posses a
$\ZZZ$-equivariant structure is that their first Chern class is
divisible by $5$ and that their second and third Chern classes are
divisible by $25$.

\section{Singular}
\label{sec:singular}

\begin{Verbatim}[fontsize=\tiny,frame=single,rulecolor=\color{red}]
LIB "random.lib";
LIB "matrix.lib";
LIB "solve.lib";

ring r = 0,(u,v,x,y,z),dp;
poly Q = u^5+v^5+x^5+y^5+z^5;

matrix f0[5][5]=u,0,0,0,0 ,0,0,0,0,y ,0,0,0,v,0 ,0,0,z,0,0 ,0,x,0,0,0;
matrix f1[5][5]=0,v,0,0,0 ,z,0,0,0,0 ,0,0,0,0,x ,0,0,0,u,0 ,0,0,y,0,0;
matrix f2[5][5]=0,0,x,0,0 ,0,u,0,0,0 ,y,0,0,0,0 ,0,0,0,0,v ,0,0,0,z,0; 
matrix f3[5][5]=0,0,0,y,0 ,0,0,v,0,0 ,0,z,0,0,0 ,x,0,0,0,0 ,0,0,0,0,u; 
matrix f4[5][5]=0,0,0,0,z ,0,0,0,x,0 ,0,0,u,0,0 ,0,y,0,0,0 ,v,0,0,0,0; 

matrix f00[5][5]=u^2,0,0,0,0,0,0,y^2,0,0,0,0,0,0,v^2,0,z^2,0,0,0,0,0,0,x^2,0;
matrix f01[5][5]=0,u*v,0,0,0,0,0,0,y*z,0,v*x,0,0,0,0,0,0,z*u,0,0,0,0,0,0,x*y;
matrix f02[5][5]=0,0,u*x,0,0,0,0,0,0,y*u,0,v*y,0,0,0,0,0,0,z*v,0,x*z,0,0,0,0;
matrix f03[5][5]=0,0,0,y*u,0,v*y,0,0,0,0,0,0,z*v,0,0,0,0,0,0,x*z,0,u*x,0,0,0;
matrix f04[5][5]=0,0,0,0,z*u,0,x*y,0,0,0,0,0,0,u*v,0,y*z,0,0,0,0,0,0,v*x,0,0;
matrix f11[5][5]=0,0,v^2,0,0,0,0,0,0,z^2,0,x^2,0,0,0,0,0,0,u^2,0,y^2,0,0,0,0;
matrix f12[5][5]=0,0,0,v*x,0,z*u,0,0,0,0,0,0,x*y,0,0,0,0,0,0,u*v,0,y*z,0,0,0;
matrix f13[5][5]=0,0,0,0,v*y,0,z*v,0,0,0,0,0,0,x*z,0,u*x,0,0,0,0,0,0,y*u,0,0;
matrix f14[5][5]=z*v,0,0,0,0,0,0,x*z,0,0,0,0,0,0,u*x,0,y*u,0,0,0,0,0,0,v*y,0;
matrix f22[5][5]=0,0,0,0,x^2,0,u^2,0,0,0,0,0,0,y^2,0,v^2,0,0,0,0,0,0,z^2,0,0;
matrix f23[5][5]=x*y,0,0,0,0,0,0,u*v,0,0,0,0,0,0,y*z,0,v*x,0,0,0,0,0,0,z*u,0;
matrix f24[5][5]=0,x*z,0,0,0,0,0,0,u*x,0,y*u,0,0,0,0,0,0,v*y,0,0,0,0,0,0,z*v;
matrix f33[5][5]=0,y^2,0,0,0,0,0,0,v^2,0,z^2,0,0,0,0,0,0,x^2,0,0,0,0,0,0,u^2;
matrix f34[5][5]=0,0,y*z,0,0,0,0,0,0,v*x,0,z*u,0,0,0,0,0,0,x*y,0,u*v,0,0,0,0;
matrix f44[5][5]=0,0,0,z^2,0,x^2,0,0,0,0,0,0,u^2,0,0,0,0,0,0,y^2,0,v^2,0,0,0;

matrix BtoC = concat( f22+f33, f4+f1+f3 );

matrix Alt2BtoBB[10*10][10*(10-1)/2]=0;
int pos=1;
for (int i=0; i<10; i++) {
    for (int j=i+1; j<10; j++) {
       Alt2BtoBB[10*i+j+1, pos] =  1;
       Alt2BtoBB[10*j+i+1, pos] = -1;
       pos++;
    }      
}      
matrix CCtoSym2C[5*(5+1)/2][5*5]=0;
int pos=1;
for (int i=0; i<5; i++) {
    for (int j=i; j<5; j++) {
       CCtoSym2C[pos, 5*i+j+1] = 1;
       CCtoSym2C[pos, 5*j+i+1] = 1;
       pos++;
    }      
}      
// BB = B tensor B is indexed by (b1,b2)	
// CB = C tensor B is indexed by  (c,b2)	
matrix BBtoCB[5*10][10*10]=0; 
for (int b1=0; b1<10; b1++) { 
  for (int b2=0; b2<10; b2++) {
    for (int c=0; c<5; c++) {
        BBtoCB[10*c+b2+1, 10*b1+b2+1] = BtoC[c+1, b1+1];
} } }
// CB = C tensor B is indexed by (c1,b)	
// CC = C tensor C is indexed by (c1,c2)	
matrix CBtoCC[5*5][5*10]=0; 
for (int c1=0; c1<5; c1++) {
  for (int b=0; b<10; b++) { 
      for (int c2=0; c2<5; c2++) {
        CBtoCC[5*c1+c2+1, 10*c1+b+1] = BtoC[c2+1, b+1];
} } }
matrix Alt2BtoCB = BBtoCB * Alt2BtoBB;
matrix CBtoSym2C = CCtoSym2C * CBtoCC;

// So far we defined polynomial matrices representing
// H^0(B) ----BtoC----> H^0(C)
// H^0(wedge^2 B) ----Alt2BtoBC----> H^0(C tensor B) ----CBtoSym2C----> H^0(Sym^2 C)

// composition must be zero
compress( CBtoSym2C * Alt2BtoCB );

// The image of BtoC lives in 5 O(3) and the quintic constraint does not matter
// The image of CBtoSym2C lives in 25 O(6), so modding out the quintic Q is important
module higgs = std(CBtoSym2C + freemodule(15)*Q);

// This computes that dim coker BtoC = 5+15+25+30 = 75
hilb(std(BtoC));

// This computes that dim coker CBtoSym2C = 6 = number of Higgs
hilb(higgs);
// Lets find the 6 representatives
kbase(higgs,6);
// Figure out the g_1 action on H^2(wedge^2 V); g_2 action is easy
reduce( u6*gen(8),     higgs);   // = g_1( z6*gen(9)    )
reduce( z6*gen(13),    higgs);   // = g_1( y6*gen(6)    )
reduce( v3yu2*gen(10), higgs);   // = g_1( u3xz2*gen(1) )
reduce( xz5*gen(15),   higgs);   // = g_1( vy5*gen(10)  )
reduce( u6*gen(1),     higgs);   // = g_1( z6*gen(13)   )
reduce( yu5*gen(6),    higgs);   // = g_1( xz5*gen(15)  )

// As a 5 x 10 matrix, BtoC has a kernel. But it requires polynomials of degree >=6
// so BtoC: H^0( Phi(1)+Phi(2) ) ---> H^0( Phi(3) ) has no kernel
module ker = std(modulo(BtoC,0*BtoC));
intvec ker_deg = 0:ncols(ker);
for (int i=1; i<=ncols(ker); i++) { ker_deg[i] = maxdeg1(ker[i]); }; ker_deg;

// similarly, Alt2BtoCB is injective
module ker = std(modulo(Alt2BtoCB,0*Alt2BtoCB));
intvec ker_deg = 0:ncols(ker);
for (int i=1; i<=ncols(ker); i++) { ker_deg[i] = maxdeg1(ker[i]); }; ker_deg;

// lets compute where the bundle is singular
ideal cym = ideal(Q);
ideal sing = cym+ideal(minor(BtoC,5,cym));
dim(std(sing)); // by homogeneity, a discrete solution set must be { u=v=x=y=z=0 }
solve(sing);    // indeed. Hence, no singularity in P^4 and V is a bundle
\end{Verbatim}

\bibliographystyle{utcaps} 
\renewcommand{\refname}{Bibliography}
\addcontentsline{toc}{section}{Bibliography} 
\bibliography{Volker}

\end{document}